# L-UNet: An LSTM Network for Remote Sensing Image Change Detection

Shuting Sun, Lin Mu, Lizhe Wang, *Fellow, IEEE*, and Peng Liu

*Abstract*— Change detection of high-resolution remote sensing images is an important task in earth observation and was extensively investigated. Recently, deep learning has shown to be very successful in plenty of remote sensing tasks. The current deep learning-based change detection method is mainly based on conventional long short-term memory (Conv-LSTM), which does not have spatial characteristics. Since change detection is a process with both spatiality and temporality, it is necessary to propose an end-to-end spatiotemporal network. To achieve this, Conv-LSTM, an extension of the Conv-LSTM structure, is introduced. Since it shares similar spatial characteristics with the convolutional layer, L-UNet, which substitutes partial convolution layers of UNet-to-Conv-LSTM and Atrous L-UNet (AL-UNet), which further using Atrous structure to multiscale spatial information is proposed. Experiments on two data sets are conducted and the proposed methods show the advantages both in quantity and quality when compared with some other methods.

## I. Introduction

LAND cover changes are frequently occurred due to human activities and natural causes. With the advent of the era of big data [1], the study of remote sensing change detection is very important to many remote sensing applications, such as vegetation, soil moisture, water surfaces, pollution, and terrain mapping.

Based on using labels or not in the process of implementation, change detections can be divided into supervised change detection and unsupervised change detection. Unsupervised change detection methods are very convenient since they do not use labels. Most of them are index methods. They suppress unchanged areas and highlight changed areas through some indexes or image algebras, such as principal component analysis (PCA) [2] or slow feature analysis (SFA) [3]. Supervised change detection methods can usually achieve higher accuracy because the labels will provide more information for them. Most of these kinds of methods are based on classification or segmentation methods, such as Bayesian classifier [4] or morphologically supervised PCA-Net [5].

In recent years, deep learning has become a very successful method in the field of machine learning [6], [7]. It can learn complex and deep features through large networks and are fastly optimized through backpropagation. It has been proved effectiveness in processing raster remote sensing data. If large artificially labeled data set in the corresponding format is provided, it is able to accomplish many vision tasks, such as accomplish scene classification, target recognition, and semantic segmentation. Change detection is to obtain the change difference map through two or more image of the same area. Essentially, most of the supervised change detection can be modeled as a problem of classification or segmentation. Therefore, a deep learning network is also adopted for change detection.

There are already some studies of change detection based on deep learning. In [8], it extracted changing features through a sparse autoencoder (SAE) and calculated the change map of each pixel through a fully convolutional network. Zhang *et al.* [9] acquired feature vector through deep belief network (DBN) and identified change area through cosine angle distance of two vectors. Wang *et al.* [10] calculated abundance map and affinity matrix of multispectral data and obtained change detection result through a full convolution network. For these kinds of methods, the core process of change analysis is carried out through traditional methods, such as log-ratio operator, cosine angle distance or abundance map, and deep learning networks only serve as classifiers or feature extraction aiding tools. Hence, we would like to further learn the multitemporal deep feature by developing a full deep learning network for change vector analysis, so as to build a real end-to-end change detection network.

At present, long short-term memory (LSTM) [11] is a powerful tool to analyze the temporal features of time sequences. There are also studies on multitemporal images, which makes use of recurrent neural network (RNN)-based structure to solve the multispectral change detection task [12]. As we know, the input data for change detection have both spatial and temporal characteristics. However, for such kind of networks, the output of LSTM layer is a $1 \times N$ dimensional vector, which would inevitably loss spatial information.

This work was supported in part by the Key- Area Research and Development Program of Guangdong Province under Grant 2020B1111020005 and in part by NSFC under Grant U1711266, Grant U2006210, Grant 41925007, Grant 61731022, and Grant 41971397. *(Corresponding author: Peng Liu.)*

Shuting Sun is with the College of Marine Science and Technology, China University of Geosciences (CUG), Wuhan 430074, China.

Lin Mu is with the College of Life Sciences and Oceanography, Shenzhen University, Shenzhen 518060, China, and also with the Southern Marine Science and Engineering Guangdong Laboratory (Guangzhou), Guangzhou 511458, China.

Lizhe Wang is with the School of Computer Science, China University of Geosciences (CUG), Wuhan 430074, China.

Peng Liu is with the Aerospace Information Research Institute, Chinese Academy of Sciences, Beijing 100094, China (e-mail: liupeng@radi.ac.cn).

In this letter, we proposed to use the convolutional LSTM (Conv-LSTM) layer [13] as the core convolutional structure, which delicately combines the convolution and recurrent structure in a single layer. Since Conv-LSTM is just a layer, we also need to nest it in a full convolution network. Based on the idea of UNet [14] which is a simple and clear end-to-end convolutional structure, we proposed to substitute the convolutional layer of UNet with Conv-LSTM to form a new architecture L-UNet. In addition, we also convert the common convolutional kernel into dilated convolutional kernel to enhance the adaptability to image offset in change detections. In the following, we will discuss how to construct as L-UNet in detail and validate its performance with different data sets.

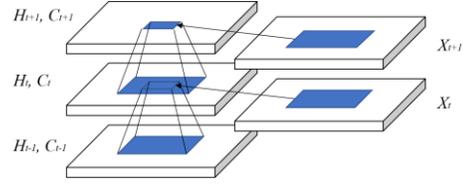

Fig. 1. Structure of Conv-LSTM layer.

## II. METHODOLOGY

### A. LSTM Layer

In LSTM, sequence data transmission and processing are realized by three control gate units: input gate, output gate, and forget gate. The structures of control gates are used for information control and make LSTM better deal with sequence data. Before addressing the recurrent structure of Conv-LSTM, we first give out fully connected LSTM (FC-LSTM), in which the weight of memory cell of previous phase is added into the calculation of memory cell. The equations of FC-LSTM are denoted as follows:

$$\begin{aligned} i_t &= \sigma(W_{xi} x_t + W_{hi} H_{t-1} + W_{ci} \odot C_{t-1} + b_i) \\ f_t &= \sigma(W_{xf} x_t + W_{hf} H_{t-1} + W_{cf} \odot C_{t-1} + b_f) \\ o_t &= \sigma(W_{xo} x_t + W_{ho} H_{t-1} + W_{co} \odot C_{t-1} + b_o) \\ C_t &= f_t \odot c_{t-1} + i_t \odot \tanh(W_{xc} x_t + W_{hc} H_{t-1} + b_c) \\ H_t &= o_t \odot \tanh(C_t). \end{aligned} \quad (1)$$

In above equations, $x_t$ is the input, $H_t$ is the output, and $C_t$ is the hidden state. $i_t$, $f_t$, and $o_t$ are the input, forget, and output gate, respectively. Each of the abovementioned parameters is related to the phase $t$, such as output cell $H_t$. Operator $\odot$ is the Hadamard product. The main feature of LSTM is its memory cell $C_t$, which accumulates state information for each phase. The cell is accessed, written, and forgotten by three control gates: input gate, output gate, and forget gate. Each of them is realized by weighted summation where $W$ (for convenient, we delete its subscript index) is the coefficient matrix that represents weight and $b$ (for convenient, we delete its subscript index) is bias of the relevant gate.

Basically, the traditional LSTM structure is to compute weighted summation of inputs of one dimension and then apply a nonlinear function to obtain an output. Without considering the 2-D spatial information, the input and output of LSTM can only be $1 \times N$ dimensional vector. It is only good at dealing with temporal features in sequence data but not able to well make full use of spatial features. However, for a change detection problem in remote sensing, both temporal features and spatial features are important. Most of the changes are not defined separately pixel by pixel but within a neighborhood by different spatial features, such as edge or texture.

### B. Conv-LSTM Layer

Considering the shortcomings of LSTM without spatial features, we extend the conventional LSTM structure into Conv-LSTM based on the idea of [13]. It holds the same recurrent structure as LSTM so that it can be used to model sequence data. The key formulas of Conv-LSTM are shown in (2), where $X_1, \ldots, X_t$ are input spatiotemporal images, they are 3-D tensors whose the last two dimensions are spatial dimensions. $C_1, \ldots, C_t$ are cell outputs, and $H_1, \ldots, H_t$ are hidden states. $i_t$, $f_t$, and $o_t$ are input, forget, and output gate, respectably. $\otimes$ denotes the convolution operator and $\odot$ denotes the Hadamard product

$$\begin{aligned} i_t &= \sigma(W_{xi} \otimes X_t + W_{hi} \otimes H_{t-1} + W_{ci} \odot C_{t-1} + b_i) \\ f_t &= \sigma(W_{xf} \otimes X_t + W_{hf} \otimes H_{t-1} + W_{cf} \odot C_{t-1} + b_f) \\ o_t &= \sigma(W_{xo} \otimes X_t + W_{ho} \otimes H_{t-1} + W_{co} \odot C_{t-1} + b_o) \\ C_t &= f_t \odot C_{t-1} + i_t \odot \tanh(W_{xc} \otimes X_t + W_{hc} \otimes H_{t-1} + b_c) \\ H_t &= o_t \odot \tanh(C_t). \end{aligned} \quad (2)$$

On account of the above equations, in time domain, recurrent units in Conv-LSTM are also weighted sum of input gate, output gate, and forget state data. However, in spatial domain, different from conventional LSTM, instead of matrix multiplication for 1-D tensor, the input-to-state and state-to-state transitions in Conv-LSTM are realized by convolution for 2-D tensor.

Concretely, in Conv-LSTM, the future state of a certain cell $C_t$ is determined by both temporal features and spatial features. The temporal feature is $f_t \odot C_{t-1}$. The spatial features are $W_{xc} \otimes X_t + W_{hc} \otimes H_{t-1} + b_c$, where inputs $X_t$ and $H_{t-1}$ will transfer their states of local neighbors to $C_t$ through convolution operator $\otimes$ in input-to-state, state-to-state, and state-to-output transitions. The convolutional layer of Conv-LSTM is explained in Fig. 1.

Consequently, Conv-LSTM and convolutional neural network (CNN) may share similar advantages in the way of learning spatial features because they all obtain the output result through a convolution kernel. For an instance, two Conv-LSTM layers can be connected to form a structure similar to full convolution network and the operation of its convolution part only affects the channel dimension so that all inputs are still independent of each other on phase channel. In addition, Conv-LSTM will have the property of Atrous convolution network if the kernel is Atrous filter.

Specifically, for remote sensing change detection problem, we define the new form of input of Conv-LSTM, e.g., $\{X_t(x, y, b)\}_{t=1}^T$, where $x$ and $y$ are the spatial index, $b$ is the band number of remote sensing images, and $t$ is the index of phases. When two Conv-LSTM layers are connected, the output of the first Conv-LSTM is $C_t(x, y, c_1)$, where $c_1$ is the number of channels, which includes hidden state value of all phases. When $C_t$ $(x, y, c_1)$ is connected to a Conv2D layer, such as the result, the output will correspondingly turn to the last hidden state, defined as $F(x, y, c_2)$, where $c_2$ is the number of categories and is also the changing mask of the result.

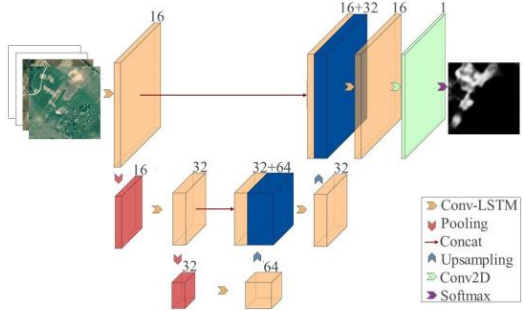

Fig. 2. Network structure of L-UNet.

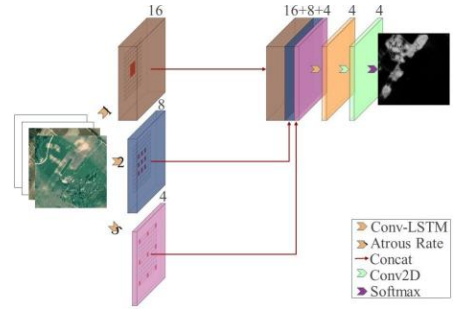

Fig. 3. Network structure of AL-UNet.

### C. Introducing Conv-LSTM Into UNet

As we know, change detection shares many similarities with image semantic segmentation. Inspired by the idea of UNet [14] which is an elegant and concise semantic segmentation model, in this letter, Conv-LSTM layers are introduced into UNet to form a new change detection framework.

The main structure of UNet consists of two parts. One is the down-sampling part. Twice identical convolutions and followed by a rectified linear unit (ReLU) and a downsampling operation with the same effect as pooling. Another is upsampling part. These upsampling layers increase the resolution of the output. In order to localize features, high-resolution features from the first part are combined with the upsampled output with the same width. Finally, the successive convolution structure can learn to assemble a precise output based on this concatenation structure.

Conventional UNet is very good at learning the spatial features in the single-phase image, but it cannot deal with multitemporal data. To make UNet adapt to multitemporal problems such as change detection, we propose that the input and partial convolution layers in the UNet structure are replaced with Conv-LSTM. Since convolution and concatenation operations are transformations in the channel dimension, poolings are transformations in the col and row dimensions, and memory function in Conv-LSTM merely occurs in the time dimension, such adjustment would still retain key features of UNet.

Moreover, if we adopt the vanilla model in UNet, both downsampling and upsampling processes contain twice consecutive identical convolutions. On the one hand, multiple convolution processes were already included within a single Conv-LSTM layer. If all convolutional layers are substituted, the network would be redundant. On the other hand, as suggested by Ronneberger et al. [14], the continuous convolution structure is more suitable to deal with image boundary problems. We design a three-layer L-UNet for change detection, which can be deeper by extending its similar U shape structure. Therefore, the twice consecutive identical convolutions structure is substituted with one Conv-LSTM layer plus a 2-D-convolution and obtain an L-UNet structure, as it is shown in Fig. 2. There is a pooling or upsampling layer between two-level Conv-LSTM layers.

### D. Improvement for L-UNet

In L-UNet, multiscale spatial information is obtained through the same upsampling and downsampling structure as UNet. According to [15], upsampling and pooling layers may unstable in reconstruction of small object, but the spatial tiering information would be a loss. The atrous convolutional layer can simultaneously play the role of convolution and pooling. As a consequence, referring to Deeplab, Atrous L-UNet (AL-UNet), a more concise architecture using Atrous spatial pyramid pooling instead of vanilla pooling structure is proposed. To avoid the gridding effect, the width of Atrous strides is 1, 2, and 5 according to the hybrid dilated convolution rule, as it is shown in Fig. 3. We simply make a summary on the proposed methods. To better learn the spatial–temporal features in remote sensing image change detection, we construct a new L-UNet structure by substituting the Conv-LSTM layer with a part of its convolutional layers. Also, in AL-Unet (a concise version of L-UNet), we further change the pooling and upsampling structure to an Atrous structure. In Section III, the proposed methods will be evaluated by different data sets.

## III. EXPERIMENTS AND RESULTS

### A. Experimental Data

In this section, two data sets (SZTAKI and Wenchuan) are used to conduct experiments.

1) SZTAKI Air Change Benchmark is a change detection data set of aerial photogrammetric images produced by Csaba Benedek. It contains 13 aerial image pairs with a size of 952 × 640 and a resolution of 1.5 m. Each image pair contains a pair of preliminary registered input images and a mask of the relevant changes, where value 1 refers to change from background to building and value 0 refers to the other situation. Sigmoid is used as a classifier, and the matching result is stretched to grayscale so that we can further metric similarity to change the category of each point. Ten image pairs of similarity in image background and ground object are selected as the training and test data sets.
2) Beichuan data set is a three-phase data set. In the 512 Wenchuan earthquake, Beichuan is the most

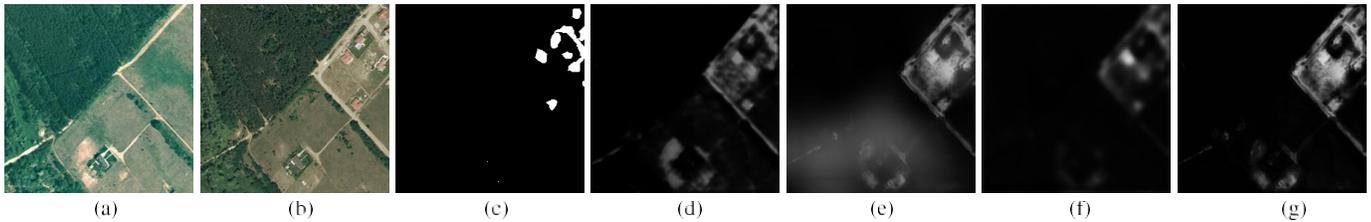

Fig. 4. Results of vanilla SZTAKI air change benchmark data set. (a) t1. (b) t2. (c) Label. (d) UNet. (e) DASNet. (f) L-UNet. (g) AL-UNet.

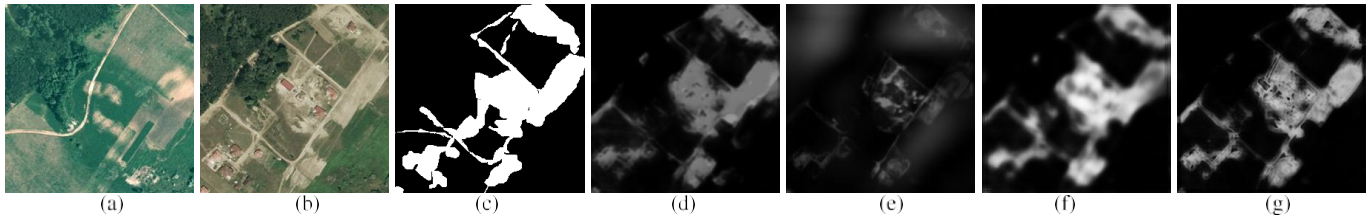

Fig. 5. Results of vanilla SZTAKI air change benchmark data set. (a) t1. (b) t2. (c) Label. (d) UNet. (e) DASNet. (f) L-UNet. (g) AL-UNet.

severely damaged area. After ten years of reconstruction work, the new Beichuan County has established many improved public facilities.

The aerial images in the years 2009, 2010, and 2011 with a resolution of 0.4 m are selected as an input, and the label is set as follows: building and background information of these images are annotated so that three binary maps are obtained. After that, relevant three binary values are concatenated into a three-digit binary number and further converted into decimal, that is, eight categories are used to represent three-phase change detection scenarios. In the results, we perform pseudo-color enhancement on output data.

### B. Result and Accuracy

In this letter, two methods are compared with the proposed L-UNet and AL-UNet. One is conventional UNet that expands the channel dimension of the input data (to 6-D in two phases and 9-D in three phases). The other is DASNet [16] method, a state-of-the-art supervised change detection method for high-resolution visible light remote sensing images. The DASNet method identifies change situation by spatial attention mechanism, which is a structure similar to conventional LSTM whose inputs are the extracting image features through CNN. In terms of accuracy, pixel accuracy and Kappa are used as an accuracy index, in the dichotomous scenario, false positive (FP, unchanged pixels that are wrongly classified as changed pixels), false negative (FN, changed pixels that are not detected), and overall error (OE, the sum of FP and FN) are also adopted.

Two groups of change detection results of original SZTAKI Air Change Benchmark are shown in Figs. 4 and 5 The images at t1 and t2 are shown in Fig. 4(a) and (b). The ground-truth label is shown in Fig. 4(c). Fig. 4(d)–(g) shows the results of UNet, DASNet, L-UNet, and AL-UNet. We can observe that the major interferences come from the bare soil areas. In Fig. 4(d) and (e), UNet and DASNet did not successfully discriminate between the building change areas and the bare soil areas. However, the proposed L-UNet in Fig. 4(f) and AL-UNet Fig. 4(g) are only lightly affected by the interferences and did not wrongly take bare soil areas as building changes.

In Fig. 5, the results of UNet in Fig. 5(d) and AL-UNet Fig. 5(g) show better performance than other methods. In Fig. 4(e), the result of DASNet shows some of the artifacts. In Fig. 4(f), L-UNet shows over smoothing.

TABLE I
ACCURACY OF VANILLA SZTAKI AIR CHANGE BENCHMARK DATA SET

| Method | UNet | DASNet | L-UNet | AL-UNet |
|---|---|---|---|---|
| Accuracy | 87.51% | 88.22% | 88.45% | 90.10% |
| Kappa | 0.7434 | 0.7540 | 0.7584 | 0.7874 |
| FP | 0.1261 | 0.1152 | 0.1114 | 0.0959 |
| FN | 0.0027 | 0.0026 | 0.0038 | 0.0031 |
| OE | 0.1288 | 0.1178 | 0.1152 | 0.0970 |

TABLE II
ACCURACY OF THREE-PHASE BEICHUAN DATA SET

| Method | UNet | L-UNet | AL-UNet |
|---|---|---|---|
| Accuracy | 80.20% | 86.13% | 88.37% |
| Kappa | 0.7458 | 0.8215 | 0.8499 |

Table I listed the precision of each method. It can be seen that L-Unet shows higher accuracy by 2%–3% than other methods. Besides, the simplified AL-UNet also provides an additional precision enhancement compared with vanilla L-UNet. In Figs. 6 and 7, there are results for different methods for the Beichuan data that are composed of three aerial images in different time phases. Quantity comparisons are listed in Table II. With the increased time phase, an LSTM structure exhibits more obvious advantages. In Table II, the accuracies of L-Unet and AL-UNet are about 5% higher than that of UNet. As mentioned earlier, because there are eight types of changes in this data set, a change difference image is more complicated. Some of the characteristics of changes are not as obvious as SZTAKI data set. However, it can still indicate from Figs. 6 and 7 that the boundaries of the changed area are more complete in the proposed L-UNet methods and the AL-UNet method.

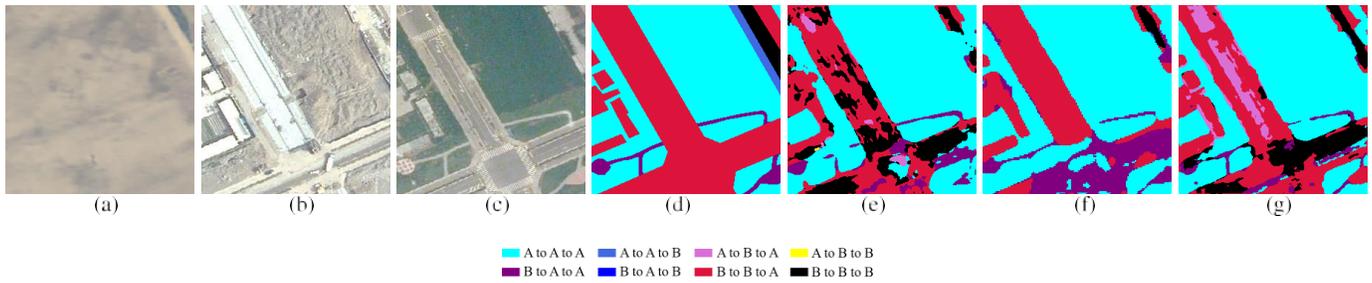

Fig. 6. Results of three-phase BeiChuan data set. In the legend, A represents the artificial object and B represents the area of background. (a) t1. (b) t2. (c) t3. (d) Truth. (e) UNet. (f) L-UNet. (g) AL-UNet.

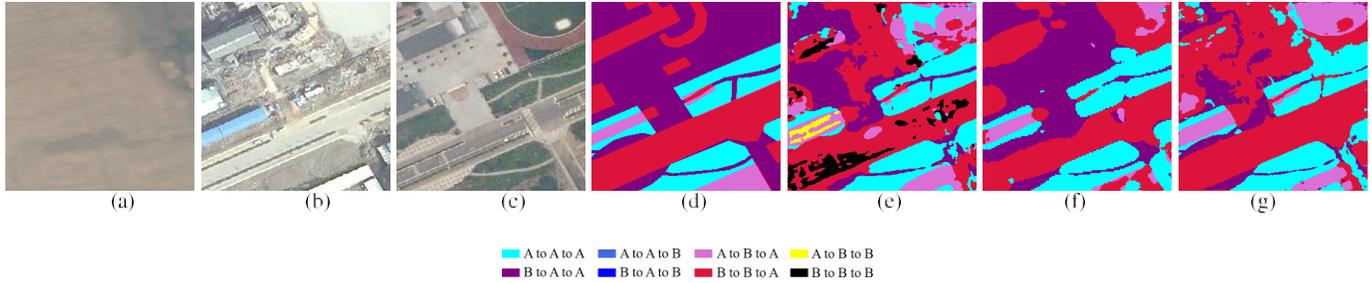

Fig. 7. Results of three-phase BeiChuan data set. In the legend, A represents the artificial object and B represents the area of background. (a) t1. (b) t2. (c) t3. (d) Truth. (e) UNet. (f) L-UNet. (g) AL-UNet.

## IV. CONCLUSION

In this letter, we proposed to construct a new end-to-end change detection deep network. Considering that the detection task is involved in both spatial and temporal features, Conv-LSTM is introduced into UNet to form L-UNet (or AL-UNet) to learn the spatiotemporal features. The proposed methods can better model the abrupt change in temporal domain and textures in spatial domain. Experiments were conducted on two data sets: SZTAKI and Beichuan. Two deep learning change detection methods (one only considered spatial-domain features and another only considered temporal-domain features) are compared with the proposed methods. The methods were tested on the data with different interferences. The results indicate the advantages of the proposed method on both quantity and quality comparisons. The accuracy of the proposed method is about 2% higher on the two phases SZTAKI data set and about 6% higher on the three-phase Beichuan data set.

## REFERENCES

[1] P. Liu, L. Di, Q. Du, and L. Wang, "Remote sensing big data: Theory, methods and applications," *Remote Sens.*, vol. 10, no. 5, p. 711, May 2018.

[2] J. S. Deng et al., "PCA-based land-use change detection and analysis using multitemporal and multisensor satellite data," *Int. J. Remote Sens.*, vol. 29, no. 16, pp. 4823–4838, 2008.

[3] C. Wu, B. Du, and L. Zhang, Slow feature analysis for change detection in multispectral imagery," *IEEE Trans. Geosci. Remote Sens.*, vol. 52, no. 5, pp. 2858–2874, May 2014.

[4] L. Bruzzone, D. F. Prieto, and S. B. Serpico, "A neural-statistical approach to multitemporal and multisource remote-sensing image classification," *IEEE Trans. Geosci. Remote Sens.*, vol. 37, no. 3, pp. 1350–1359, May 1999.

[5] R. Wang, J. Zhang, J. Chen, L. Jiao, and M. Wang, "Imbalanced learning-based automatic SAR images change detection by morphologically supervised PCA-net," *IEEE Geosci. Remote Sens. Lett.*, vol. 16, no. 4, pp. 554–558, Apr. 2019, doi: 10.1109/LGRS.2018.2878420.

[6] P. Liu, H. Zhang, and K. B. Eom, "Active deep learning for classification of hyperspectral images," *IEEE J. Sel. Topics Appl. Earth Observ. Remote Sens.*, vol. 10, no. 2, pp. 712–724, Feb. 2017.

[7] P. Liu, K.-K.-R. Choo, L. Wang, and F. Huang, "SVM or deep learning? A comparative study on remote sensing image classification," *Soft Comput.*, vol. 21, no. 23, pp. 7053–7065, Dec. 2017.

[8] M. Gong, H. Yang, and P. Zhang, "Feature learning and change feature classification based on deep learning for ternary change detection in SAR images," *ISPRS J. Photogramm. Remote Sens.*, vol. 129, pp. 212–225, Jul. 2017.

[9] H. Zhang, M. Gong, P. Zhang, L. Su, and J. Shi, "Feature-level change detection using deep representation and feature change analysis for multispectral imagery," *IEEE Geosci. Remote Sens. Lett.*, vol. 13, no. 11, pp. 1666–1670, Nov. 2016.

[10] Q. Wang, Z. Yuan, Q. Du, and X. Li, "GETNET: A general End-to-End 2-D CNN framework for hyperspectral image change detection," *IEEE Trans. Geosci. Remote Sens.*, vol. 57, no. 1, pp. 3–13, Jan. 2019.

[11] S. Hochreiter and J. Schmidhuber, "Long short-term memory," *Neural Comput.*, vol. 9, no. 8, pp. 1735–1780, 1997.

[12] L. Mou, L. Bruzzone, and X. X. Zhu, "Learning Spectral-Spatial-Temporal features via a recurrent convolutional neural network for change detection in multispectral imagery," *IEEE Trans. Geosci. Remote Sens.*, vol. 57, no. 2, pp. 924–935, Feb. 2019.

[13] S. H. I. Xingjian et al., "Convolutional LSTM network: A machine learning approach for precipitation nowcasting," in *Proc. Adv. Neural Inf. Process. Syst.*, 2015, pp. 802–810.

[14] O. Ronneberger, P. Fischer, and T. Brox, "U-Net: Convolutional networks for biomedical image segmentation," in *Proc. Int. Conf. Med. Image Comput. Comput.-Assist. Intervent.* Cham, Switzerland: Springer, 2015, pp. 234–241.

[15] P. Wang et al., "Understanding convolution for semantic segmentation," in *Proc. IEEE Winter Conf. Appl. Comput. Vis. (WACV)*, Mar. 2018, pp. 1451–1460.

[16] J. Chen et al., "DASNet: Dual attentive fully convolutional siamese networks for change detection of high resolution satellite images," 2020, arXiv:2003.03608. [Online]. Available: http://arxiv.org/abs/2003.03608